\documentclass{emulateapj}

\usepackage{amsmath}
\usepackage{aas_macros}
\usepackage{natbib}

\newcommand{\brg}{Br$\gamma \:$}

\newcommand\tna{\,\tablenotemark{a}}
\newcommand\tnb{\,\tablenotemark{b}}
\newcommand\tnc{\,\tablenotemark{c}}
\newcommand\tnd{\,\tablenotemark{d}}
\newcommand\tne{\,\tablenotemark{e}}
\newcommand\tnf{\,\tablenotemark{f}}

\newcommand{\kms}{km s$^{-1}$}
\newcommand{\msun}{M$_{\odot}$ }

\newcommand{\mcomp}{$M_{\text{comp}}$}
\newcommand{\mcompsini}{$M_{\text{comp}}\sin{i}$}
\newcommand{\vtick}{\textsc{\char13}}
\newcommand{\mtot}{$M_{\text{tot}}$}

\usepackage{hyperref}
\extrafloats{100}

\shorttitle{Is S0-2 a spectroscopic binary?}
\shortauthors{CHU ET AL.}

\begin{document}

\title{Investigating the Binarity of S0-2: Implications for its Origins and Robustness as a Probe of the Laws of Gravity around a Supermassive Black Hole}

\author{\sc Devin S. Chu\altaffilmark{1}, Tuan Do\altaffilmark{1}, Aurelien Hees\altaffilmark{1}, Andrea Ghez\altaffilmark{1}, Smadar Naoz\altaffilmark{1}, Gunther Witzel\altaffilmark{1}, Shoko Sakai\altaffilmark{1}, Samantha Chappell\altaffilmark{1}, Abhimat K. Gautam\altaffilmark{1}, Jessica R. Lu\altaffilmark{2}, Keith Matthews\altaffilmark{3}}
\altaffiltext{1}{Department of Physics and Astronomy, University of California, Los Angeles, CA 90095, USA}
\altaffiltext{2}{Astronomy Department, University of California, Berkeley, CA 94720, USA}
\altaffiltext{3}{Division of Physics, Mathematics, and Astronomy, California Institute of Technology, Pasadena, CA 91125, USA}

\begin{abstract}
The star S0-2, which orbits the supermassive black hole (SMBH) in our Galaxy with a period of 16 years, provides the strongest constraint on both the mass of the SMBH and the distance to the Galactic center. S0-2 will soon provide the first measurement of relativistic effects near a SMBH. We report the first limits on the binarity of S0-2 from radial velocity monitoring, which has implications for both understanding its origin and robustness as a probe of the central gravitational field.  With 87 radial velocity measurements, which include 12 new observations presented, we have the data set to look for radial velocity variations from S0-2's orbital model. Using a Lomb-Scargle analysis and orbit fitting for potential binaries, we detect no radial velocity variation beyond S0-2's orbital motion and do not find any significant periodic signal. The lack of a binary companion does not currently distinguish between different formation scenarios for S0-2. The upper limit on the mass of a companion star (\mcomp) still allowed by our results has a median upper limit of \mcomp \ $\sin i \leq$ 1.6 \msun for periods between 1 and 150 days, the longest period to avoid tidal break up of the binary. We also investigate the impact of the remaining allowed binary system on the measurement of the relativistic redshift at S0-2's closest approach in 2018. While binary star systems are important to consider for this experiment, we find plausible binaries for S0-2 will not alter a 5$\sigma$ detection of the relativistic redshift.


\end{abstract}

\keywords{Galaxy: center --- stars: early-type --- techniques: high angular resolution --- gravitation}

\section{Introduction}    \label{sect:intro}

The source S0-2 is one of the most well studied stars at the Galactic center. It is important for our understanding of the properties of the Galaxy\vtick s central potential. In particular, it has provided the proof of the existence of a supermassive black hole (SMBH), the characterization of the SMBH properties (mass and distance) and the laws of gravitation \citep{Ghez:2005ck, Ghez:2008ty, Gillessen:2009fn,Gillessen:2017fa, Boehle:2016ko, hees:2017ab}. S0-2 is also notable because spectroscopic studies have revealed that it, along with most of the ``S-stars'' located within 1\arcsec \ of the black hole, is a young main-sequence B star \citep{Ghez:2003iw,Eisenhauer:2005gh,Martins:2008im,Habibi:2017}. This discovery raised questions about their formation mechanism, since traditional star formation would be disrupted by the tidal forces of the black hole \citep{Morris:1993fp}. 

Works have investigated ways the S-stars may have formed and how these stars relate to the rest of the nuclear star cluster. Many theories for the S-stars\vtick \ formation have been proposed \citep[see][for a review]{Alexander:2005ij}. Two general classifications of mechanisms are considered for the S-stars: (1) binary star systems scattered from outside the region and then tidally disrupted, leaving behind one component of the original binary while the other is ejected as a hypervelocity star \citep{Hills:1988br,Perets:2007fo}, and (2) S-stars formed in the clockwise disk and then migrated to the central arcsecond around the SMBH \citep{Levin:2007ez,Lockmann:2008be,Merritt:2009ab}. Previous works have also investigated how these S-stars relate to the clockwise disk, Wolf-Rayet stars, G2-like sources and evolved giants in the region \citep{Paumard:2006fo,Lu:2009es,Bartko:2009hh,Do:2009kz,Do:2013fn,Madigan:2014fp,Chen:2014ck,Witzel:2014kv,Witzel:2017vk,Phifer:2013bm}.

With binary stars playing a leading role in many of the S-star formation and evolution scenarios as well as in scenarios of other Galactic center stars, observational searches for binaries are important. Thus far, photometric variations has been the primary search method. Several binaries have been revealed \citep{Ott:1999gn,Martins:2006,Rafelski:2007gu,Pfuhl:2014eba}, although none among the S-stars. However, eclipsing binaries are expected to be only a small fraction of the true binary population. With radial velocity (RV) measurements that now span more than a decade, there is an opportunity to search for RV variations in the brightest S-star cluster members.

Furthermore, S0-2 will be at its closest approach to the SMBH in 2018, which provides the opportunity to measure the relativistic redshift in S0-2\vtick s RV \citep{zucker:2006fk,angelil:2010qd,angelil:2011lq,hees:2017ab}. This first direct observation of a relativistic effect on S-stars orbit will improve with time after 2018 and be followed by other relativistic measurement such as the advance of the periastron.  If S0-2 is actually a spectroscopic binary, it will bias the relativistic redshift measurement if binarity is not considered.

In this work, we explore the possibility of S0-2 to be a spectroscopic binary. This paper is organized as follows: In Section \ref{sect:data} we describe the observations and data used in this work, including new RV measurements. Section \ref{sec:binary} describes the search for a companion star and the characterization of allowed hidden companions. Section \ref{sec:redshift} describes the impact allowed spectroscopic binaries would have on the relativistic redshift measurement. Section \ref{sect:discuss} interprets the results of the analysis and implications for S0-2 being a single star and the robustness of gravitational redshift measurements and future relativity studies based on S0-2\vtick s orbital motion.

\section{Observations and Data}    \label{sect:data}

This investigation includes previously reported astrometric and spectroscopic data, as well as new spectroscopic data taken with the W. M. Keck Observatory (WMKO). All the WMKO spectroscopic observations used for S0-2 RV measurements are summarized in Table~\ref{tab:newobs}.

\subsection{S0-2 Radial Velocities} \label{sect:spec}

\subsubsection{Previously Reported Data}\label{sec:previousRV}

Over the past 16 years, S0-2 has been closely monitored spectroscopically. In the published literature, 24 RV measurements beginning in the year 2000 have been reported from WMKO \citep{Ghez:2003iw,Ghez:2005ck,Ghez:2008ty,Boehle:2016ko} and 40 measurements beginning in 2003 from the VLT \citep{Eisenhauer:2005gh,Gillessen:2009fg,Gillessen:2017fa,Habibi:2017}. Many of the RV measurements are based on multiple nights of observations. For this analysis, we are interested in the presence of binaries, which for S0-2 can have periods as short as $\sim$1 day. We therefore reextract S0-2\vtick s spectra from the previously calibrated WMKO data on a nightly basis for the following nights which were previously combined: 2009 May 5 and 6 to 2009.334, 2010 May 5 and 8 to 2010.349, 2012 June 8-11 to 2012.441, 2012 July 21 and 22 to 2012.556, 2012 August 12 and 13 to 2012.616, 2013 May 14 and 16 to 2013.369, 2013 July 25-27 to 2013.566, and 2013 August 10-13 to 2013.612 (see Table~\ref{tab:newobs}). This increases the Keck data set to 38 points for this time period. For the VLT, 7 out of 41 epochs are reported to be derived from multiple nights of data.

\subsubsection{New Spectroscopic Data}

We report new spectroscopic observations for S0-2 obtained using the integral field spectrograph OSIRIS \citep{Larkin:2006} on the W. M. Keck I telescope with the laser guide star AO system. These data were observed between 2014 to 2016. Details about the filters and integration time relate to these observations are given in Table~\ref{tab:newobs}.   The RV observations and data analysis follow the same procedures used for earlier WMKO S0-2 RV measurements~\citep{Ghez:2008ty,Do:2013fn}. The 8 new RV measurements, along with the RV measurements from Section~\ref{sec:previousRV} (38 Keck  and 41 VLT) result in 87 total RV measurements used in this work (see Table \ref{tab:S0-2_measure}).
 
\begin{deluxetable*}{lcccccc}
\tablecolumns{7} 
\tablewidth{0pc} 
\tablecaption{Summary of Keck Spectroscopic Observations}
\tablehead{ 
    \multicolumn{3}{c}{Date}  & 
	\colhead{$N_{\text{frames}} \times t_{\text{int}}$} &
	\colhead{FWHM\tna} &
	\colhead{Filter} &
	\colhead{Scale} \\
	\cline{1-3} \\
    \colhead{(UT)} &
	\colhead{(MJD)} &
    \colhead{(Epoch)} &
	\colhead{(s)} &
	\colhead{(mas)} &
	\colhead{} &
	\colhead{(mas)}
}
\startdata
2000-06-23		&  	51718.50	&  	2000.476	&  	36  $\times$ 300	&		   	&	$K$\tnb &	18 	\\
2002-06-02		&  	52427.50	&  	2002.418	&  	7  $\times$ 1200	&		  	&	$K$\tnc	&	20 	\\
2002-06-03		&  	52428.50	&  	2002.420	&  	4  $\times$ 1200	&		  	&	$K$\arcmin \tnc	&	20 	\\
2003-06-08		&  	52798.50	&  	2003.433	&  	2  $\times$ 1200	&		  	&	$K$\arcmin \tnc	&	20 	\\
2004-06-23		&  	53179.50	&  	2004.476	&  	16  $\times$ 1200	&		  	&	$K$\arcmin \tnc	&	20 	\\
2005-05-30		&  	53520.50	&  	2005.410	&  	7  $\times$ 1200	&		  	&	$K$\arcmin \tnc	&	20	\\
2005-07-03		&  	53554.50	&  	2005.503	&  	7  $\times$ 900   	&	58	  	&	Kbb 	&	20	\\
2006-05-23		&  	53878.50	&  	2006.390	&  	4  $\times$ 900 	&	74	  	&	Kbb 	&	35	\\
2006-06-18		&  	53904.50	&  	2006.461	&  	9  $\times$ 900   	&	65	  	&	Kn3 	&	35	\\
2006-06-30		&  	53916.50	&  	2006.494	&  	9  $\times$ 900   	&	59	  	&	Kn3 	&	35	\\
2006-07-01		&  	53917.50	&  	2006.497	&  	9  $\times$ 900   	&	64	  	&	Kn3 	&	35	\\
2007-05-21		&  	54241.50	&  	2007.384	&  	2 $\times$ 900   	&	86		&	Kn3 	&	35	\\
2007-07-19		&  	54300.29	&  	2007.545	&  	2 $\times$ 900   	&	56	  	&	Kn3 	&	35	\\
2008-05-16		&  	54602.50	&  	2008.372	&  	11 $\times$ 900   	&	57	  	&	Kn3 	&	35	\\
2008-07-25		&  	54672.28	&  	2008.563	&  	9  $\times$ 900 	&	60	  	&	Kn3 	&	35	\\
2009-05-05		&  	54956.50	&  	2009.342	&  	7   $\times$ 900	&	60	  	&	Kn3 	&	35	\\
2009-05-06		&  	54957.50	&  	2009.344	&  	12  $\times$ 900	&	69	  	&	Kn3 	&	35	\\
2010-05-05		&  	55321.50	&  	2010.341	&  	6  $\times$ 900 	&	67	  	&	Kn3 	&	35	\\
2010-05-08		&  	55324.50	&  	2010.349	&  	11  $\times$ 900 	&	69	  	&	Kn3 	&	35	\\
2011-07-10		&  	55752.33	&  	2011.520	&  	6   $\times$ 900   	&	71	  	&	Kn3 	&	35	\\
2012-06-08		&  	56086.50	&  	2012.435	&	4  	$\times$ 900	&	87	  	&	Kn3 	&	35	\\
2012-06-09		&  	56087.50	&  	2012.438	&	3  	$\times$ 900	&	66	  	&	Kn3 	&	35	\\
2012-06-11		&  	56089.50	&  	2012.444	&  	7  	$\times$ 900	&	64	  	&	Kn3 	&	20	\\
2012-07-21		&  	56129.31	&  	2012.553	&  	3   $\times$ 900  	&	77	  	&	Kn3 	&	35	\\
2012-07-22		&  	56130.31	&  	2012.555	&  	7   $\times$ 900   	&	81	 	&	Kn3 	&	35	\\
2012-08-12		&  	56151.33	&  	2012.613	&  	6   $\times$ 900   	&	56	  	&	Kn3 	&	35	\\
2012-08-13		&  	56152.27	&  	2012.615	&  	7   $\times$ 900	&	99	  	&	Kn3 	&	35	\\
2013-05-11		&  	56423.50	&  	2013.358	&  	11  $\times$ 900	&	73	  	&	Kbb		&	35	\\
2013-05-12		&  	56424.50	&  	2013.361	&  	11  $\times$ 900  	&	62	  	&	Kbb 	&	35	\\
2013-05-13		&  	56425.50	&  	2013.363	&  	12  $\times$ 900	&	61	  	&	Kbb	 	&	35	\\
2013-05-14		&  	56426.50	&  	2013.366	&  	11  $\times$ 900   	&	61	  	&	Kn3	 	&	35	\\
2013-05-16		&  	56428.50	&  	2013.372	&  	7  $\times$ 900		&	98	  	&	Kn3	 	&	20	\\
2013-05-17		&  	56429.50	&  	2013.374	&  	7  $\times$ 900 	&	64   	&	Kn3	 	&	20	\\
2013-07-25		&  	56498.33	&  	2013.563	&  	11  $\times$ 900 	&	79   	&	Kn3	 	&	35	\\
2013-07-26		&  	56499.34	&  	2013.566	&  	6  $\times$ 900 	&	73   	&	Kn3	 	&	35	\\
2013-07-27		&  	56500.33	&  	2013.568	&  	11  $\times$ 900 	&	66   	&	Kn3	 	&	35	\\
2013-08-10		&  	56514.29	&  	2013.607	&  	7 $\times$ 900		&	62   	&	Kn3 	&	35	\\		
2013-08-11		&  	56515.31	&  	2013.609	&  	9 $\times$ 900		&	69   	&	Kn3 	&	35	\\
2013-08-13		&  	56517.29	&  	2013.615	&  	12 $\times$ 900		&	67   	&	Kn3 	&	35	\\
2014-05-18		&	56795.50	& 	2014.376	& 	13 $\times$ 900		& 	66 		&	Kn3		&	35	\\
2014-05-23		&	56800.50	& 	2014.390	& 	10 $\times$ 900		& 	76 		&	Kn3		&	35	\\
2014-07-03		&	56841.36	& 	2014.502	& 	8 $\times$ 900		& 	66 		&	Kn3		&	35	\\
2015-05-04		&	57146.50	& 	2015.337	& 	5 $\times$ 900		& 	68 		&	Kn3		&	35	\\
2015-07-21		&	57224.35	& 	2015.551	& 	5 $\times$ 900		& 	56 		&	Kn3		&	35	\\
2016-05-14		&	57522.50	& 	2016.367	& 	8 $\times$ 900		& 	78 		&	Kbb		&	35	\\
2016-05-15		&	57523.50	& 	2016.370	& 	4 $\times$ 900		& 	80 		&	Kbb		&	35	\\
2016-05-16		&	57524.50	& 	2016.372	& 	8 $\times$ 900		& 	84 		&	Kbb		&	35	\\

\enddata
\tablenotetext{a}{Average FWHM of S0-2 in the mosaic made of all frames, measured by fitting a two-dimensional Gaussian to the source.}
\tablenotetext{b}{Taken with NIRSPEC slit spectrograph}
\tablenotetext{c}{Taken with NIRC2 slit spectrograph}
\label{tab:newobs}
\end{deluxetable*}

\subsection{Characteristics of the Two Datasets}
The Keck and VLT data sets are analyzed in a similar manner and appear to be consistent with one another. The two datasets are analyzed with the same standard spectroscopic calibration procedures and the absolute wavelength solutions are both determined from the OH sky emmision lines. The radial velocity of S0-2 is measured from both data sets by fitting a Gaussian to the \brg absorption line. The reported average RV uncertainties are very similar,  33 \kms and 45 \kms for Keck and VLT, respectively. Furthermore, for the 4 Keck and VLT points taken within 10 days of each other, 3 of the points were within 1$\sigma$ of each other. The one exception is the Keck 2003.433 point, which differs from a nearby VLT point by 2$\sigma$.  We conclude there is no significant systematic difference.

\subsection{Removing S0-2\vtick s Long-term RV Variations} \label{sect:longterm_remove}

\begin{figure}[htb]
\centering
\includegraphics[width=\linewidth]{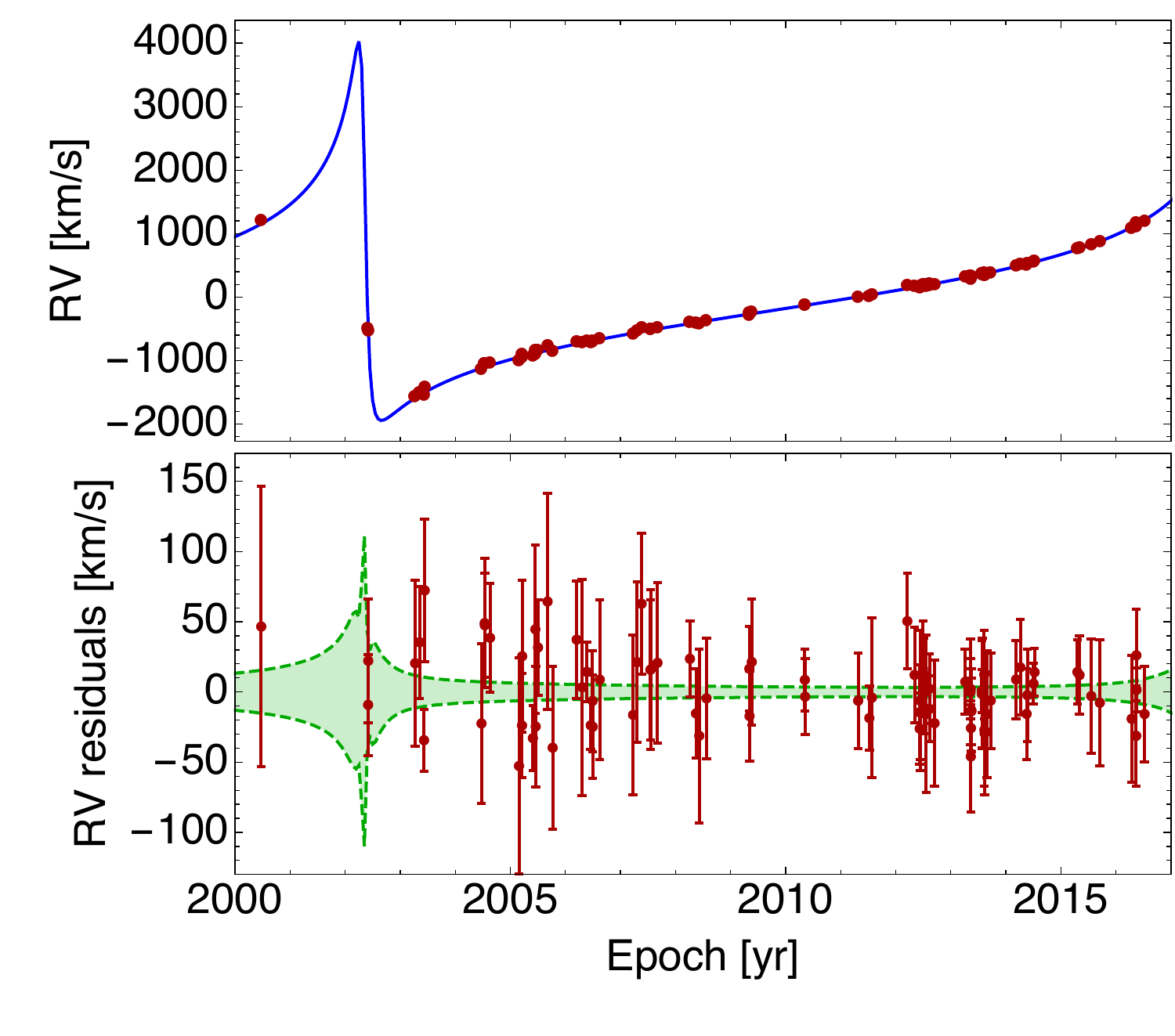}
\caption{
\small
Top: S0-2\vtick s radial velocity measurements over time and best fit model. Bottom: Residual radial velocity curve. Dashed lines are the 1$\sigma$ model uncertainties.
\label{fig:S0-2_rv}
}
\end{figure}

Before searching for short-term RV variations, we remove the long-term RV variations from S0-2\vtick s orbital motion around the SMBH. To create the long-term RV model, a simultaneous orbital fit of S0-2 and S0-38 was performed using the same S0-2 and S0-38 astrometry and process as \citet{Boehle:2016ko}, but with the S0-2 RVs in Table \ref{tab:S0-2_measure} and S0-38 RVs from \citet{Boehle:2016ko,Gillessen:2017fa}. One additional change is the format of time used. In this work, we use Modified Julian Date (MJD). The reported time is the approximate average time of the observations taken during the night. For convenience, we also report the Universal Time (UT) and epoch time reported in Julian years of 365.25 days since J2000. Previously, \citet{Boehle:2016ko} defined the epoch year as 365.24 days. The orbital parameters resulting from the fit are consistent with \citet{Boehle:2016ko} within 1$\sigma$ and are presented in the Appendix~\ref{sect:full_orbit_res}. The RV data, model and residual are shown in Figure \ref{fig:S0-2_rv} and given in Table \ref{tab:S0-2_measure}. The average scatter around the orbit residual is 20 \kms \ with a standard deviation of 26 \kms. The Keck and VLT datasets are individually consistent with these values.

\section{Is S0-2 a Binary?}    \label{sec:binary}

In this section, we use two different methods to to search for periodic signals in the RV residuals: (i) the Lomb-Scargle periodogram \citep{Lomb:1976, Scargle:1982eu, 2013Astropy} and (ii)  a Bayesian fit for potential binaries. The former method provides quick overview of the data;  we note that while it may not be as effective at detecting highly eccentric binaries (e.g. phase dispersion measure \citep{Stellingwerf:1972}, minimum string length \citep{Dworetsky:1983}), it is an computationally efficient method for detecting periodic signals in unevenly spaced data (see Appendix~\ref{sect:lombscargle}). The latter method provides a more complete and robust, albeit more computationally expensive approach and allows us to derive upper limits on the orbital parameters of an hypothetical binary companion to S0-2.

We can place an upper limit on the orbital period of any possible companion around S0-2 of 119.2 days based on a binary disruption criteria. A binary would be tidally disrupted at closest approach to the SMBH if it has a separation greater than the Hill radius ($r_{\text{H}}$). The Hill radius is given by
\begin{equation}
	r_{\text{H}} = a_\textrm{S0-2} (1 - e_\textrm{S0-2})\sqrt[3]{\frac{M_{\text{Primary}}}{3M_{BH}}}\, ,
	\label{eq:hill}
\end{equation}
where $M_{\text{Primary}}$ is the primary mass, $a_\textrm{S0-2}$ and $e_\textrm{S0-2}$ are the semimajor axis and eccentricity for the binary-black hole system (the corresponding values have been derived from the orbital fit presented in Section~\ref{sect:longterm_remove} whose result is presented in Appendix~\ref{sect:full_orbit_res}). This Hill Radius limit is a conservative limit since any eccentricity of the inner binary system would decrease the stability of the system. We take the condition $a(1+e) < r_{\text{H}}$, where $e$ is the eccentricity and $a$ the semi-major axis of the inner binary, to allow for longterm stability \citep{Naoz:2016}, which leads to the following constraint on the binary period $P$
\begin{equation}
	P^{2} < \frac{4\pi^{2}}{3GM_{BH}}a_\textrm{S0-2}^{3}\frac{(1 - e_\textrm{S0-2})^{3}}{(1+e)^3}\frac{M_{\text{Primary}}}{M_{tot}}\, .
	\label{eq:hill_limit}
\end{equation}
This condition needs to be fulfilled to avoid a disruption of the binary. We therefore sampled periods between 1 and 150 days to search for a significant periodic signal. 

\begin{figure}[htb]
\centering
\includegraphics[width=\linewidth]{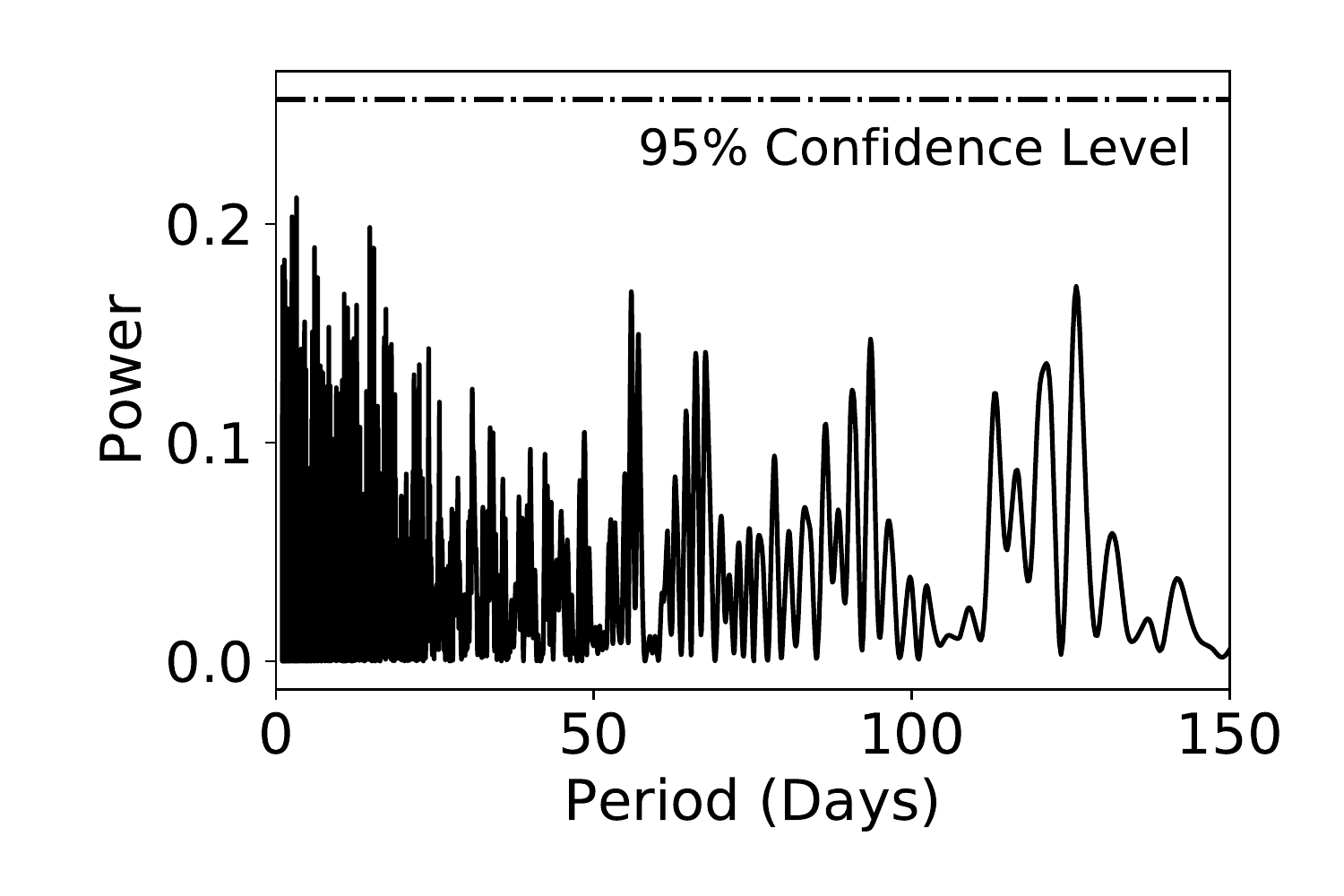}
\caption{
\small
Lomb-Scargle periodogram of S0-2\vtick s residual RV curve. The black dash-dotted line is the 95\% confidence level detection value. No power reaches the 95\% confidence level detection value implying that no significant periodic signal is found in the observations. 
\label{fig:LS_S02}
}
\end{figure}

The Lomb-Scargle analysis on S0-2\vtick s RV does not reveal any statistically significant peak (see Figure~\ref{fig:LS_S02}). We note that the structure of this periodogram is unaffected by the model uncertainties over the period range searched. In order to determine the relationship between periodogram power and statistical significance, we ran a series of Monte Carlo simulations. We first generated 100,000 simulated residual RV curves with no periodic signal. The simulated points had the same observation times and uncertainties as the data and were drawn from a Gaussian centered around 0 \kms. We produced a periodogram for each simulated RV curve and found the maximum peak power value. We then looked at the distribution of these maximum power values and made a cumulative distribution function (CDF). We used this CDF to determine the significance for periodic detections. These simulations set the 95\% confidence level detection limit to be 0.25, shown in a dotted line in Figure \ref{fig:LS_S02}. The periodogram corresponding to S0-2\vtick s observations never reaches this value, which implies that no significant periodic signal is found in the current data and that observations are consistent with a single star model.

\begin{figure}[htb]
	\centering
	\includegraphics[width=\linewidth]{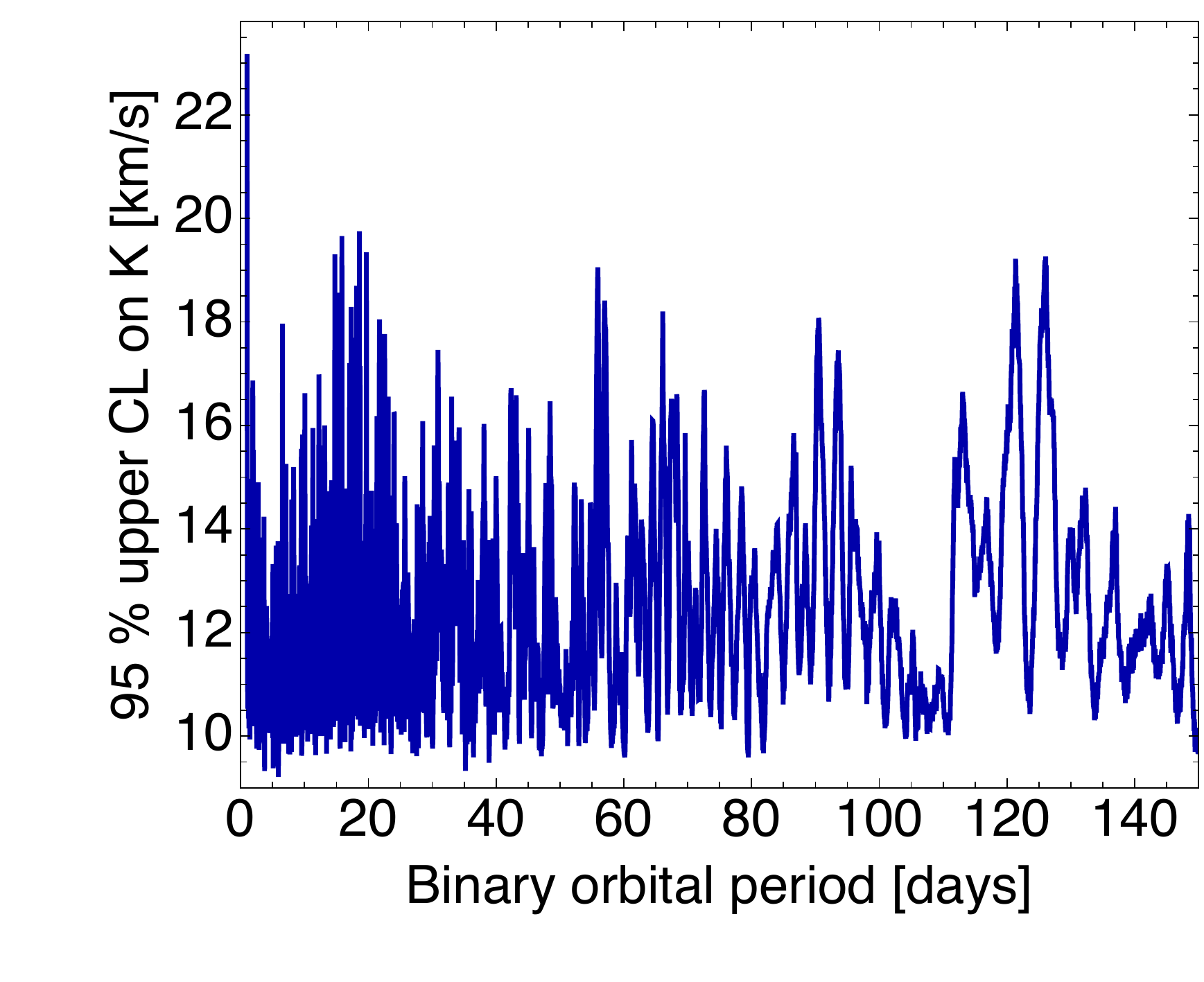}
 	\caption{
 	\small  Upper limits on the amplitude of RV variations induced by possible undetected companion stars. Plotted at the 95\% upper confidence limit, the amplitude of RV variations that could be induced by a hidden binary system ($K$ as defined by equations~\ref{eq:eccentric_rv} and \ref{eq:Kdef}) as a function of the binary orbital period.
 	\label{fig:periodvmax}
 }
\end{figure}

Since no evidence of a binary for S0-2 is found, we can place an upper limit on the amplitude on the RV variations induced by a binary system. In order to infer such a limit, we fit the S0-2 RV residuals with a binary star RV model plus a constant. The following equation was used to model the RV curve of an eccentric binary system~\citep{Hilditch:2001aa}
\begin{equation}
	RV = K \frac{\sqrt{1 - e^{2}}\cos{E}\cos{\omega} - \sin{E}\sin{\omega}}{1 - e \cos{E}}\, ,
	\label{eq:eccentric_rv}
\end{equation}
with
\begin{equation}
	K = \frac{2\pi a\sin{i}}{P}\, ,
	\label{eq:Kdef}
\end{equation}
and where $e$ is the binary eccentricity, $\omega$ the argument of periastron, $E$ the eccentric anomaly determined by solving the Kepler equation, $i$ the inclination, $P$ the period  and $a$ the semimajor axis. This model is parametrized using the following 5 variables: the offset $O$, the RV amplitude $K$, the eccentricity $e$, the argument of periastron $\omega$ and the mean longitude at J2000 (noted $L_0$). The use of the mean longitude at J2000 is preferred to the usual time of closest approach which is not bounded and not defined in case of circular orbits~\citep{Hilditch:2001aa}. For different fixed binary orbital periods $P$, we fitted this model to the RV residuals using a {\sc MultiNest} sampler~\citep{Feroz:2008fi,Feroz:2009aa,Feroz:2013aa}.  The resulting 95\% upper confidence limit on $K$ for $\sim$3000 orbital periods ranging from 1 to 150 days is shown in Figure~\ref{fig:periodvmax}.

Like the Lomb-Scargle analysis, this method is also sensitive to periodic signals in the data. A periodic signal would yield a significant non-zero peak in the value of $K$ in the posterior, as opposed to a power law decreasing posterior. This method also takes inherently into account the other orbital parameters of the binary system that can affect the shape of the curve (e.g. eccentricity).

\begin{figure*}[htb]
\centering
	\includegraphics[width=.75\linewidth]{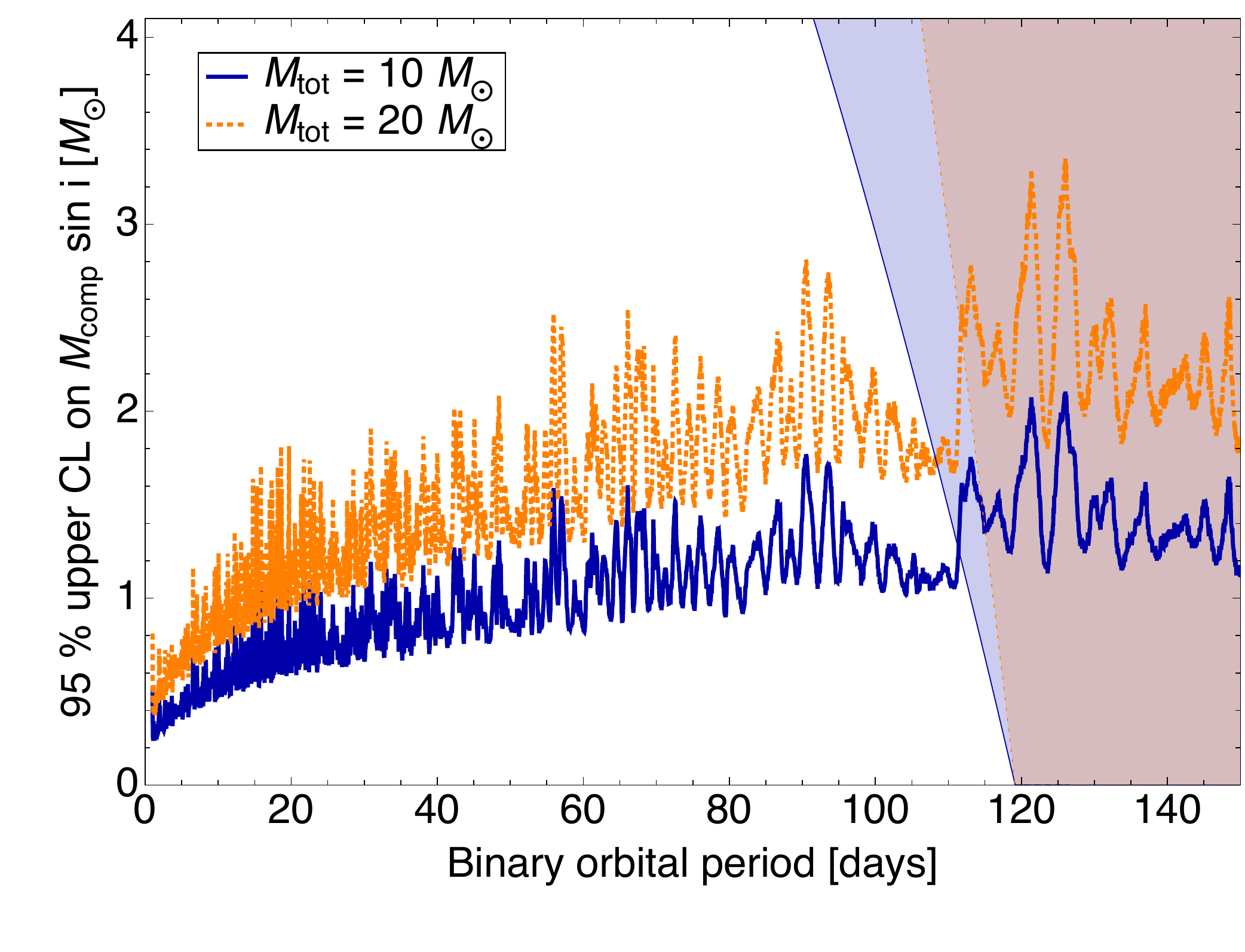}
\caption{
\small Upper limits on allowed companion star masses. Plotted are the 95\% upper confidence limits on  \mcompsini \ as a function of the binary orbital period assuming two different values for the total mass of the binary system. The shaded regions represent the area excluded by the Hill radius limits from equation \ref{eq:hill_limit} while adopting a circular orbit (which corresponds to the most conservative limit).
\label{fig:periodmass}
}
\end{figure*}

With one further assumption, this analysis also allows us to constrain the mass of a hypothetical companion. Assuming a total mass of the system (\mtot), we transform  the sampling (i.e. the chain) resulting from the {\sc MultiNest} run into a companion mass limit by using
\begin{equation}
	M_{\text{comp}}\sin{i}  = \left(\frac{PM_{tot}^{2}}{2\pi G}\right)^{1/3} \ K\, ,
	\label{eq:Binary_mass}
\end{equation}
where $M_\text{comp}$ is the companion mass and $i$ the inclination of the binary system. From this transformed chain, we can derive an upper 95\% confidence limit on $M_{\text{comp}}\sin{i}$.  This limit depends on the total mass \mtot \ used in Equation \ref{eq:Binary_mass}. In this work, two extreme values for \mtot \ are considered: (i) a low value of \mtot = 10 \msun (\citet{Habibi:2017} reported the mass of S0-2 as $13.6^{+2.2}_{-1.8}$ \msun) and (ii) a high value of \mtot = 20 \msun. The upper limit on \mcomp \ $\sin{i}$ is shown in Figure \ref{fig:periodmass} as well as the excluded region inferred by theoretical arguments based on the on the binary disruption criteria\footnote{Interactions from background stars \citep{Hopman:2009bp} and the eccentric Kozai-Lidov mechanism \citep{Li:2017} can also disrupt or merge binaries. We do not consider these scenarios because these effects depend on many variables, such as the age of the binary.} and characterized by equation \ref{eq:hill_limit}.  The median upper 95\% confidence limit for \mcompsini \ for all periods is 1.6 \msun assuming a total mass of 20 \msun while its maximal value is 2.8 \msun at 93.5 days period. These values decrease by 36 \% for a total mass \mtot~of 10.0 \msun.

\section{Impact of Hidden Allowed Companions On Measurement of S0-2\vtick s Relativistic Redshift}    \label{sec:redshift}

As anticipated by many theorists, observations of short-period stars orbiting the SMBH in our Galactic center are currently opening a new window to test the gravitational theory and to measure relativistic effects (see e.g.  \citet{rubilar:2001dq,zucker:2006fk,will:2008fk,borka:2013nx,zakharov:2016xy,johannsen:2016sh,johannsen:2016jh,psaltis:2016jk,hees:2017aa,Iorio:2017} and references therein). The relativistic redshift on S0-2's RV is the first relativistic effect expected to be detected with S0-2\vtick s closest approach in 2018 \citep{zucker:2006fk,angelil:2010qd,angelil:2011lq,hees:2017ab}. This measurement of the relativistic redshift will improve with time in the future and will also be followed by measurements of more relativistic effects such as the advance of S0-2's periastron.  In the case where S0-2 is a binary system, the measurement of the relativistic effects like the redshift would be altered. The goal of this section is to quantify the impact of a binarity of S0-2 on the measurement of its relativistic redshift.

One way to measure the relativistic redshift is to model the total RV as $RV = \left[RV \right]_\textrm{Newton} + \Upsilon \left[RV \right]_\textrm{rel}$, where $\left[RV \right]_\textrm{Newton}$ is the standard Newtonian RV, $\Upsilon$  a dimensionless parameter whose value is equal to 1 in GR and $\left[ RV \right]_\textrm{rel}$ is the first order relativistic contribution to the RV given by
\begin{equation}
	\left[ RV \right]_\textrm{rel} = \frac{v^2}{2c}+\frac{GM_{BH}}{rc}\, ,
\end{equation}
with $c$ the speed of light in a vacuum and $r$ the norm of the star's position with respect to the SMBH and $v$ the norm of its velocity. The first term is a contribution due to special relativity while the second term corresponds to the gravitational redshift. For a Keplerian orbit, the two contributions are exactly the same (up to a constant factor), meaning that only their combination can be measured. The relativistic redshift contribution to S0-2's RV reaches 200 km/s at closest approach in 2018 while the Newtonian part is ranging from -2000 km/s to 4000 km/s (see Figure~\ref{fig:S0-2_rv}).  The idea is to fit $\Upsilon$ simultaneously with the other parameters in the orbital fit: a value significantly different from 0 but compatible with 1 would be a successful detection of the relativistic redshift while a value significantly different from 1 would indicate a deviation from GR. The goal of this section is to quantify the impact of a plausible binary for S0-2 on the determination of $\Upsilon$.

The methodology consists in simulating data assuming S0-2 is a binary star using a relativistic modeling (in particular we use $\Upsilon=1$) and analyze these data using a modeling where S0-2 is a single star and where $\Upsilon$ is a free parameter. The deviation $\Upsilon-1$ obtained in this analysis is therefore entirely due to the fact that S0-2 has been simulated as a binary star.

More precisely, we simulate astrometric and RV data for S0-2 using a relativistic modeling that includes the R\"omer time delay and the redshift (see e.g. \cite{Alexander:2005ij}). The simulated epochs correspond to epochs where we actually have data (see  Table \ref{tab:S0-2_measure}) and for each simulated data, we assign an uncertainty that corresponds to the actual measurement. In addition to existing data, we included simulated data for 2018: 10 spectroscopic observations which were assigned an uncertainty of 25 km/s and 4 astrometric observations which were assigned an uncertainty of 0.3 mas. The epochs for these additional observations have been chosen to optimize a redshift measurement within the 2018 observation window. At this step, the simulated data corresponds to perfect measurements in the case where S0-2 is a single star. Therefore, an orbital fit using these simulated data recovers the input value, i.e. gives an estimate of $\Upsilon=1$ as expected. 

To these simulated data, we then add the signature produced by a binary star given by Eq. (\ref{eq:eccentric_rv}). The obtained data now corresponds to a binary system. This dataset is then used in a one star orbital fit that includes the $GM_{BH}$, the distance to our Galactic center $R_0$, the position and velocity of the BH, the 6 orbital  parameters for the star and the relativistic redshift parameter $\Upsilon$. The impact of the binarity of S0-2 on the redshift measurement will be given by the estimated value\footnote{We use the median as the estimated value from the {\sc MultiNest} fit.} of $\Upsilon-1$.  

\begin{figure*}[ht]
\centering
  \includegraphics[scale=0.45]{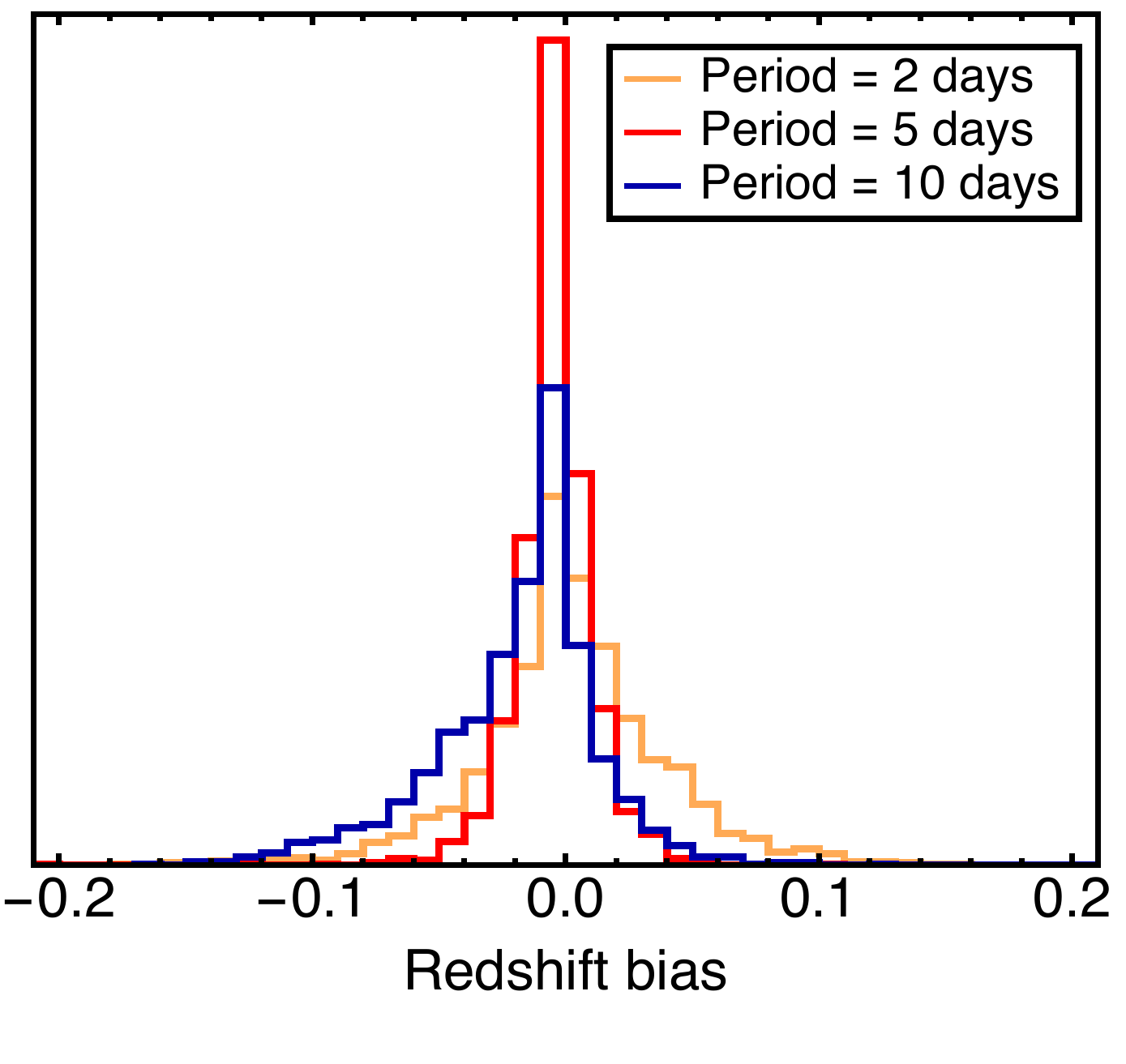}\hspace*{2cm} 
  \includegraphics[scale=0.45]{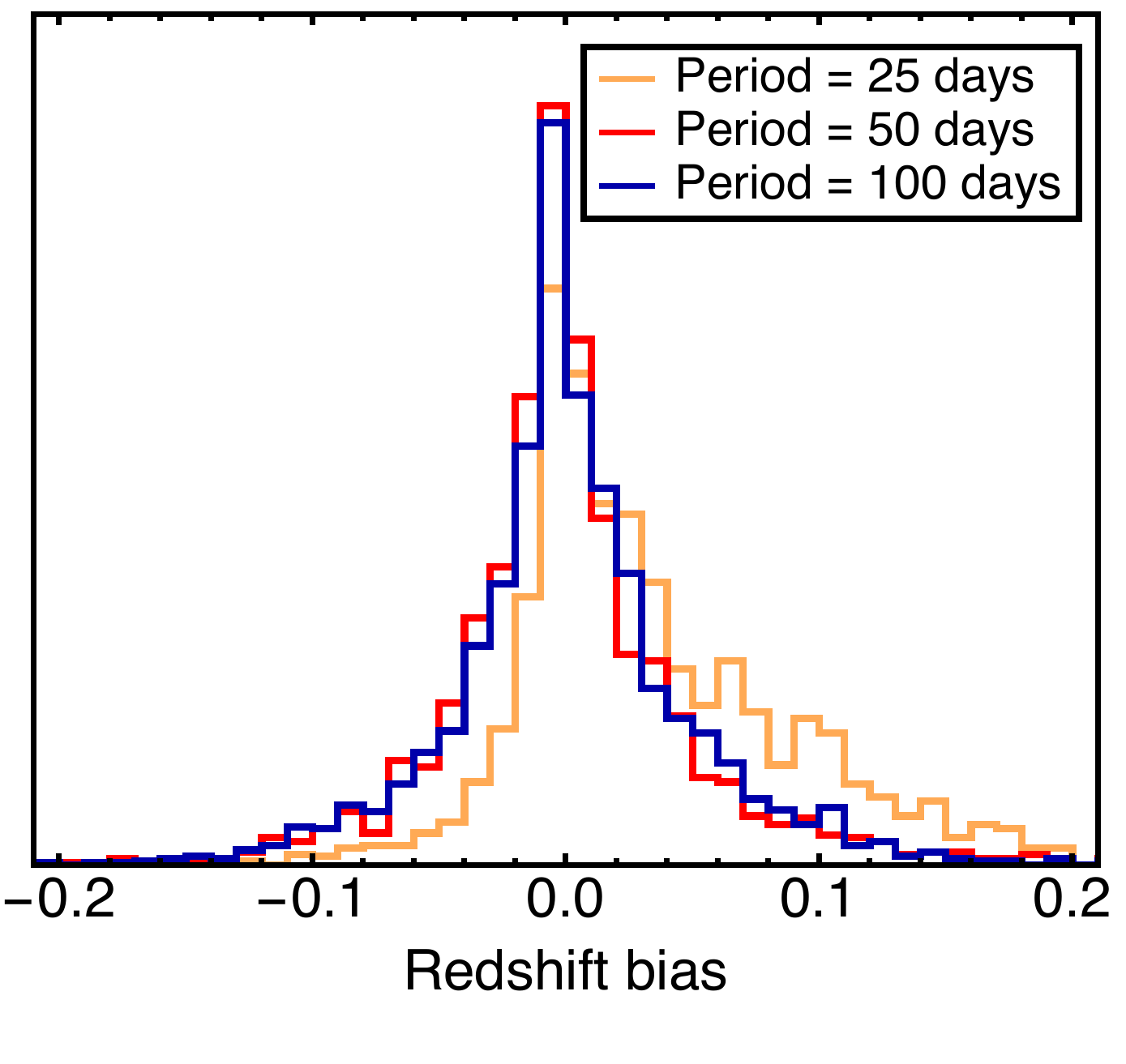}
\caption{
\small
Bias on the estimation of relativistic redshift ($\Upsilon-1$). Plotted are the imposed biases from model fits using simulated S0-2 observations, which extend through 2018 and which include RV variations induced by binary stars systems allowed by the current S0-2 RV data.  The different curves correspond to different binary periods. In general, the periods that have stricter companion limits, result in smaller biases; the median binary amplitude limits correspond to a redshift bias of 0.03.
\label{fig:redshift_bias}
}
\end{figure*}
\begin{deluxetable}{lcc}[ht]
	\tablecolumns{2} 
	\tablewidth{0pc} 
	\tablecaption{Impact of a binarity of S0-2 on a measurement of the relativistic redshift.}
	\tablehead{    	\colhead{Binary period} & \colhead{95 \% C.L. on} &	\colhead{Uncertainty on the redshift} \\
	& \colhead{$K$ [km/s]\tna} 	& \colhead{due to the binarity ($\sigma_\Upsilon$)}  }
	\startdata
	2	days & 16.0  & 0.031 	\\
	5	days & 10.4  & 0.011 	\\
	10  days & 12.2  & 0.026     \\
	25  days & 12.6  & 0.051 	\\
	50  days & 11.0  & 0.036 	\\
	100 days & 12.9  & 0.039
\enddata
\tablenotetext{a}{$K$ is the amplitude of the RV variations induced by a binary as introduced in equations~\ref{eq:eccentric_rv} and~\ref{eq:Kdef}. The 95 \% upper confidence limits on $K$ are presented on Figure~\ref{fig:periodvmax}.}
\label{tab:redshift}
\end{deluxetable}

This procedure has been performed for 6 different binary orbital periods: 2, 5, 10, 25, 50 and 100 days. For each of these periods, we draw 2000 samples for the other binary orbital parameters (eccentricity $e$, RV amplitude $K$, argument of periastron $\omega$ and longitude at J2000 $L_0$) from the posterior probability distribution function of the fit described in Section~\ref{sec:binary}. The resulting distributions for the bias in the redshift measurement $\Upsilon-1$ are presented in Fig.~\ref{fig:redshift_bias}. Furthermore, the values of the half of the 68\% upper limit on the absolute value of the redshift bias are given in Tab.~\ref{tab:redshift}. In general, the periods that have stricter limits on the amplitude of the RV variations ($K$) induced by a hidden binary, result in smaller biases; the median binary amplitude limits correspond to a redshift bias of 0.03.

Although a plausible binary for S0-2 can bias the measurement of the relativistic redshift, this bias is always smaller than the uncertainty corresponding to a 5$\sigma$ detection of the redshift (a 5$\sigma$ detection is characterized by $\sigma_\Upsilon=0.2$).

\section{Discussion and Conclusion}    \label{sect:discuss}
\subsection{S0-2 in its astrophysical context} \label{subsect:star_form}
This is the first work that investigates S0-2 as a spectroscopic binary. Previous searches have concentrated on brighter sources such as IRS16SW and E60 \citep{Pfuhl:2014eba}, located beyond the S-star cluster region (outside of $\sim$0.04 pc). This work has pushed the RV searches to 2 magnitudes fainter from  K = 12.0 (E60) down to K = 14.0; physically, this magnitude difference corresponds to the difference between evolved Wolf-Rayet stars and main-squence B stars. The improvements here are driven by the large number of RV measurements available for S0-2 from the long-baseline monitoring programs for this source \citep[e.g.,][]{Boehle:2016ko,Gillessen:2017fa}.

While we detect no significant binary signal in the RV variations, we have been able to place stringent limits on the companion mass. Our limits of 1.6 \msun \ are consistent with other observations of the star. For example, given our 95\% confidence limit of 1.6 M$_\odot$, a star would have an observed brightness of K $\sim$ 18 mag at the Galactic center, corresponding to a factor of 40 times less flux than S0-2. This brightness ratio is consistent with the fact that S0-2's spectrum shows no sign of another set of spectral features, even with 10 years of spectra combined \citep{Habibi:2017}. While our mass limits has a $\sin i$ degeneracy, the lack of detection of a double-lined source also shows that a face-on binary system is also very unlikely.

The lack of a binary companion does not distinguish between different formation scenarios for S0-2 at this time. No companion is expected if S0-2 is the remaining companion of a hypervelocity star \citep[e.g.][]{Hills:1988br,Yu:2003,Perets:2009}, or if it is the product of a merger \citep[e.g.][]{Sana:2012,Phifer:2013bm,Witzel:2014kv,Stephan:2016eh}. Scattering from the young star cluster (at 0.1 to 0.5 pc) could also bring S0-2 in without a companion \citep{Perets:2007fo,Madigan:2014fp}. While the current observations of S0-2 are unable to distinguish these scenarios, companion searches of the other S-stars should be able to provide a much more comprehensive test of the formation scenarios for the S-stars. We have concentrated on S0-2 because the other S-stars are all fainter, which results in lower precision in their RV compared to S0-2. Additional measurements should improve the sensitivity of companion searches.

\subsection{S0-2 as a new probe of fundamental physics} \label{subsect:rel_discuss}

The relativistic redshift at S0-2\vtick s closest approach in 2018 will be the first measurement of its kind, so understanding all sources of bias will be especially important for a significant detection. We have shown that a binary companion below our current detection limit for S0-2 can bias this measurement, as shown in Sec.~\ref{sec:redshift}. Nevertheless, this bias is always smaller than the uncertainty corresponding to a 5$\sigma$ detection of the redshift. The values reported in Tab.~\ref{tab:redshift} should be taken into account in the estimation of the uncertainty produced by all systematic effects in 2018. We would like to emphasize that the 2018 observations campaign is expected to reduce this possible bias.

Continued monitoring of S0-2 beyond 2018 provides further opportunities to observe other relativistic effects, such as the advance of the periastron. The impact that a plausible binary system would have on these relativistic measurements is left for further work.

\section{Acknowledgements}	\label{sect:acknowledge}

We are grateful for the helpful and constructive comments received from the our referee. We thank M.~R.~Morris and E.~E.~Becklin for their comments and long term efforts on the Galactic Center Orbits Initiative. We also thank the staff of the Keck Observatory, especially Jim Lyke, Randy Campbell, Gary Puniwai, Heather Hershey, Hien Tran, Scott Dahm, Jason McIlroy, Joel Hicock, and Terry Stickel for all their help in obtaining the new observations. Support for this work at UCLA was provided by NSF grant AST-1412615, the W. M. Keck Foundation for support of the \textit{NStarsOrbits} Project, the Levine-Leichtman Family Foundation, Ken and Eileen Kaplan Student Support Fund, the Preston Family Graduate Fellowship (held by D.C. and A.G.), the Galactic Center Board of Advisors and the Janet Marott Student Travel Awards. The W. M. Keck Observatory is operated as a scientific partnership among the California Institute of Technology, the University of California, and the National Aeronautics and Space Administration. The authors wish to recognize that the summit of Maunakea has always held a very significant cultural role for the indigenous Hawaiian community. We are most fortunate to have the opportunity to observe from this mountain. The Observatory was made possible by the generous financial support of the W. M. Keck Foundation.

\bibliographystyle{apj}
\bibliography{apj-jour,myrefs}

\LongTables
\begin{deluxetable*}{lllrrrrrrrc}
\tablecolumns{11} 
\tablewidth{0pc} 
\tablecaption{S0-2 Radial Velocity Measurements}
\tablehead{ 
    \colhead{UT} & 
    \colhead{MJD} &
	\colhead{Epoch\tna} &
	\colhead{$RV_\textrm{obs}$} &
	\colhead{$RV$ $\sigma$} &
	\colhead{$V_\textrm{LSR}$\tnb} &
	\colhead{$RV_\textrm{LSR}$\tnc} &
	\colhead{$RV$\tnd} &
	\colhead{Model} &
	\colhead{Model $\sigma$} &
	\colhead{Residual\tne} \\
    \colhead{} & 
    \colhead{} &
	\colhead{} &
	\colhead{(km s$^{-1}$)} &
	\colhead{(km s$^{-1}$)} &
	\colhead{(km s$^{-1}$)} &
	\colhead{(km s$^{-1}$)} &
	\colhead{Source}		&
	\colhead{(km s$^{-1}$)} &
	\colhead{(km s$^{-1}$)} &
	\colhead{(km s$^{-1}$)}
}
\startdata

2000-06-23	&  	51718.50	&  	2000.476	&  	1192   	&	100   	&	7 		&	1199  	&	(2)		&     1152   &      15  &      47    \\
2002-06-02	&  	52427.50	&  	2002.418	&  	-513   	&	36   	&	18 		&	-495  	&	(2)		&     -486   &      28  &      -9    \\
2002-06-03	&  	52428.50	&  	2002.420	&  	-550   	&	44   	&	18 		&	-532  	&	(2)		&     -554   &      27  &      22    \\
2003-04-09	&  	52739.23	&  	2003.271	&  			&	59 		&			&	-1571  	&	(3)		&    -1592   &      12  &      21    \\
2003-05-09	&  	52769.18	&  	2003.353	&  			&	40 		&			&	-1512  	&	(3)		&    -1547   &      11  &      35    \\
2003-06-08	&  	52798.50	&  	2003.433	&  	-1556   &	22   	&	15 		&	-1541  	&	(2)		&    -1507   &      10  &     -34    \\
2003-06-12	&  	52803.15	&  	2003.446	&  			&	51 		&			&	-1428  	&	(3)		&    -1500   &      10  &      72    \\
2004-06-23	&  	53179.50	&  	2004.476	&  	-1151   &	57   	&	8 		&	-1143  	&	(2)		&    -1121   &       7  &     -22    \\
2004-07-14	&  	53200.91	&  	2004.535	&  			&	46 		&			&	-1055  	&	(3)		&    -1104   &       7  &      49    \\
2004-07-15	&  	53201.64	&  	2004.537	&  			&	37 		&			&	-1056  	&	(3)		&    -1104   &       7  &      48    \\
2004-08-18	&  	53236.34	&  	2004.632\tnf&  			&	39 		&			&	-1039  	&	(3)		&    -1078   &       7  &      39    \\
2005-02-26	&  	53428.46	&  	2005.158	&  			&	77 		&			&	-1001  	&	(3)		&     -948   &       6  &     -53    \\
2005-03-18	&  	53448.18	&  	2005.212	&  			&	37 		&			&	-960  	&	(3)		&     -936   &       6  &     -24    \\
2005-03-19	&  	53449.28	&  	2005.215	&  			&	54 		&			&	-910  	&	(3)		&     -935   &       6  &      25    \\
2005-05-30	&  	53520.50	&  	2005.410	&  	-945   	&	23   	&	19 		&	-926  	&	(2)		&     -893   &       6  &     -33    \\
2005-06-15	&  	53536.94	&  	2005.455	&  			&	60 		&			&	-839  	&	(3)		&     -884   &       6  &      45    \\
2005-06-17	&  	53539.13	&  	2005.461	&  			&	43 		&			&	-907  	&	(3)		&     -882   &       6  &     -25    \\
2005-07-03	&  	53554.50	&  	2005.503	&  	-845   	&	34   	&	3 		&	-842  	&	(2)		&     -874   &       6  &      32    \\
2005-09-04	&  	53618.02	&  	2005.677\tnf&  			&	77 		&			&	-774  	&	(3)		&     -838   &       6  &      64    \\
2005-10-08	&  	53651.63	&  	2005.769\tnf&  			&	58 		&			&	-860  	&	(3)		&     -820   &       6  &     -40    \\
2006-03-15	&  	53810.51	&  	2006.204	&  			&	42 		&			&	-702  	&	(3)		&     -739   &       5  &      37    \\
2006-04-21	&  	53847.40	&  	2006.305	&  			&	77 		&			&	-718  	&	(3)		&     -721   &       5  &       3    \\
2006-05-23	&  	53878.50	&  	2006.390	&  	-715   	&	21   	&	23 		&	-692  	&	(2)		&     -707   &       5  &      14    \\
2006-06-18	&  	53904.50	&  	2006.461	&  	-728   	&	17   	&	10 		&	-718  	&	(2)		&     -694   &       5  &     -24    \\
2006-06-30	&  	53916.50	&  	2006.494	&  	-699   	&	36   	&	4 		&	-695  	&	(2)		&     -689   &       5  &      -6    \\
2006-07-01	&  	53917.50	&  	2006.497	&  	-717   	&	37   	&	4 		&	-713  	&	(2)		&     -688   &       5  &     -25    \\
2006-08-16	&  	53963.92	&  	2006.624\tnf&  			&	57 		&			&	-658  	&	(3)		&     -667   &       5  &       9    \\
2007-03-25	&  	54185.26	&  	2007.230	&  			&	57 		&			&	-586  	&	(3)		&     -570   &       5  &     -16    \\
2007-04-21	&  	54212.29	&  	2007.304	&  			&	57 		&			&	-537  	&	(3)		&     -558   &       5  &      21    \\
2007-05-21	&  	54241.50	&  	2007.384	&  	-507   	&	50   	&	24 		&	-483  	&	(2)		&     -546   &       5  &      63    \\
2007-07-19	&  	54300.29	&  	2007.545	&  	-502   	&	50   	&	-4 		&	-506  	&	(2)		&     -522   &       5  &      16    \\
2007-07-20	&  	54302.14	&  	2007.550\tnf&  			&	57 		&			&	-505  	&	(3)		&     -521   &       5  &      16    \\
2007-09-03	&  	54347.06	&  	2007.673\tnf&  			&	57 		&			&	-482  	&	(3)		&     -503   &       4  &      21    \\
2008-04-05	&  	54562.20	&  	2008.262\tnf&  			&	27 		&			&	-394  	&	(3)		&     -418   &       4  &      24    \\
2008-05-16	&  	54602.50	&  	2008.372	&  	-443   	&	32   	&	26 		&	-417  	&	(2)		&     -402   &       4  &     -15    \\
2008-06-06	&  	54623.92	&  	2008.431	&  			&	62 		&			&	-425  	&	(3)		&     -394   &       4  &     -31    \\
2008-07-25	&  	54672.28	&  	2008.563	&  	-373   	&	43   	&	-7 		&	-380  	&	(2)		&     -375   &       4  &      -5    \\
2009-05-05	&  	54956.50	&  	2009.342	&  	-282   	&	30   	&	30 		&	-252  	&	(2)		&     -268   &       4  &      17    \\
2009-05-06	&  	54957.50	&  	2009.344	&  	-315   	&	32   	&	30 		&	-285  	&	(2)		&     -268   &       4  &     -17    \\
2009-05-20	&  	54972.37	&  	2009.385	&  			&	45 		&			&	-241  	&	(3)		&     -262   &       4  &      21    \\
2010-05-08	&  	55324.50	&  	2010.349	&  	-152   	&	22   	&	29 		&	-123  	&	(2)		&     -131   &       4  &       9    \\
2010-05-09	&  	55326.30	&  	2010.354	&  			&	27 		&			&	-134  	&	(3)		&     -131   &       4  &      -3    \\
2011-04-26	&  	55678.03	&  	2011.317	&  			&	34 		&			&	-3  	&	(3)		&        3   &       3  &      -6    \\
2011-07-10	&  	55752.33	&  	2011.520	&  	14   	&	23   	&	0 		&	14  	&	(2)		&       32   &       3  &     -19    \\
2011-07-26	&  	55769.35	&  	2011.567	&  			&	57 		&			&	35  	&	(3)		&       39   &       3  &      -4    \\
2012-03-17	&  	56004.20	&  	2012.210	&  			&	34 		&			&	185  	&	(3)		&      135   &       3  &      50    \\
2012-05-04	&  	56052.42	&  	2012.342	&  			&	34 		&			&	167  	&	(3)		&      155   &       3  &      12    \\
2012-06-08	&	56086.50	&	2012.435	&	128		&	25		&	15		&	143		&	(2)		&      169   &       3  &     -26    \\
2012-06-09	&	56087.46	&	2012.438	&	141		&	34		&	14		&	155		&	(2)		&      170   &       3  &     -15    \\
2012-06-11	&  	56089.50	&  	2012.444	&  	151   	&	50   	&	13 		&	164  	&	(2)		&      171   &       3  &      -6    \\
2012-06-29	&  	56107.93	&  	2012.494	&  			&	34 		&			&	195  	&	(3)		&      179   &       3  &      16    \\
2012-07-06	&  	56114.87	&  	2012.513	&  			&	34 		&			&	186  	&	(3)		&      182   &       3  &       4    \\
2012-07-21	&  	56129.31	&  	2012.553	&  	178   	&	56   	&	-6 		&	172  	&	(2)		&      188   &       3  &     -16    \\
2012-07-22	&  	56130.31	&  	2012.555	&  	200   	&	8   	&	-6 		&	194  	&	(2)		&      188   &       3  &       6    \\
2012-08-12	&  	56151.33	&  	2012.613	&  	213   	&	25   	&	-13 	&	200  	&	(2)		&      197   &       3  &       2    \\
2012-08-13	&  	56152.27	&  	2012.615	&  	199   	&	24   	&	-14 	&	186  	&	(2)		&      198   &       3  &     -12    \\
2012-09-14	&  	56185.00	&  	2012.705	&  			&	45 		&			&	190  	&	(3)		&      212   &       3  &     -22    \\
2013-04-05	&  	56388.45	&  	2013.262	&  			&	23 		&			&	313  	&	(3)		&      306   &       3  &       7    \\
2013-05-11	&  	56423.50	&  	2013.358	&  	295   	&	38   	&	28 		&	323  	&	(1)		&      323   &       3  &      -0    \\
2013-05-12	&  	56424.50	&  	2013.361	&  	298   	&	22   	&	27 		&	325  	&	(1)		&      323   &       3  &       2    \\
2013-05-13	&  	56425.50	&  	2013.363	&  	251   	&	39   	&	27	 	&	278  	&	(1)		&      324   &       3  &     -46    \\
2013-05-14	&  	56426.50	&  	2013.366	&  	272   	&	16   	&	27	 	&	298  	&	(2)		&      324   &       3  &     -26    \\
2013-05-16	&  	56428.50	&  	2013.372	&  	286   	&	24   	&	26	 	&	311  	&	(2)		&      325   &       3  &     -14    \\ 
2013-05-17	&  	56429.50	&  	2013.374	&  	287   	&	30   	&	25	 	&	312  	&	(1)		&      326   &       3  &     -13    \\
2013-07-25	&	56498.33	&	2013.563	&	367		&	15		&	-7		&	360		&	(2)		&      360   &       4  &       1    \\
2013-07-26	&	56499.34	&	2013.566	&	366		&	39		&	-7		&	359		&	(2)		&      360   &       4  &      -2    \\
2013-07-27	&  	56500.33	&  	2013.568	&  	367   	&	39   	&	-8	 	&	360  	&	(2)		&      361   &       4  &      -1    \\ 
2013-08-10	&	56514.29	&	2013.607	&	393		&	32		&	-13		&	380		&	(2)		&      368   &       4  &      12    \\
2013-08-11	&	56515.31	&	2013.609	&	354		&	40		&	-13		&	341		&	(2)		&      368   &       4  &     -27    \\
2013-08-13	&  	56517.29	&  	2013.615	&  	353   	&	44   	&	-14 	&	340  	&	(2)		&      369   &       4  &     -30    \\
2013-08-27	&  	56531.99	&  	2013.655	&  			&	45 		&			&	361  	&	(3)		&      377   &       4  &     -16    \\
2013-09-22	&  	56557.92	&  	2013.726	&  			&	34 		&			&	384  	&	(3)		&      390   &       4  &      -6    \\
2014-03-08	&  	56725.57	&  	2014.185	&  			&	28 		&			&	490  	&	(3)		&      481   &       4  &       9    \\
2014-04-06	&  	56754.06	&  	2014.263	&  			&	34 		&			&	515  	&	(3)		&      497   &       4  &      18    \\
2014-05-17	&  	56795.50	&  	2014.376	&  	481   	&	32   	&	25	 	&	506  	&	(1)		&      522   &       4  &     -16    \\
2014-05-22	&  	56800.50	&  	2014.390	&  	500   	&	33   	&	23 		&	523  	&	(1)		&      525   &       4  &      -2    \\
2014-07-02	&  	56841.36	&  	2014.502	&  	553   	&	15   	&	3 		&	556  	&	(1)		&      550   &       4  &       6    \\
2014-07-09	&  	56848.30	&  	2014.521	&  			&	17 		&			&	568  	&	(3)		&      554   &       4  &      14    \\
2015-04-19	&  	57132.46	&  	2015.299	&  			&	23 		&			&	765  	&	(3)		&      751   &       5  &      14    \\
2015-05-03	&  	57146.50	&  	2015.337	&  	743   	&	28   	&	31	 	&	774  	&	(1)		&      762   &       5  &      12    \\
2015-07-20	&  	57224.35	&  	2015.551	&  	829   	&	41   	&	-5	 	&	823  	&	(1)		&      826   &       5  &      -3    \\
2015-09-15	&  	57281.12	&  	2015.706	&  			&	45 		&			&	869  	&	(3)		&      877   &       5  &      -8    \\
2016-04-13	&  	57492.23	&  	2016.284	&  			&	45 		&			&	1081  	&	(3)		&     1100   &       8  &     -19    \\
2016-05-14	&  	57522.50	&  	2016.367	&  	1081   	&	36   	&	26	 	&	1107  	&	(1)		&     1138   &       8  &     -31    \\
2016-05-15	&  	57523.50	&  	2016.370	&  	1139   	&	33   	&	26 		&	1165  	&	(1)		&     1140   &       8  &      26    \\
2016-05-16	&  	57524.50	&  	2016.372	&  	1117   	&	16   	&	26	 	&	1142  	&	(1)		&     1141   &       8  &       1    \\
2016-07-08	&  	57578.06	&  	2016.519	&  			&	34 		&			&	1198  	&	(3)		&     1214   &       9  &     -16    
                                             	
\enddata
\tablenotetext{a}{The epoch time is reported in Julian years 365.25 days since J2000.}
\tablenotetext{b}{The values came from $rvcorrect$ task in IRAF, with an error less than 1 \kms \citep{Kerr:1986}. For ease of viewing, values have been rounded.}
\tablenotetext{c}{Velocity after applying the $V_\textrm{LSR}$ correction. For ease of viewing, values have been rounded.}
\tablenotetext{d}{Measurement reported in (1) this work, (2) \citet{Boehle:2016ko} and (3) \citet{Gillessen:2017fa}.}
\tablenotetext{e}{$RV_\textrm{LSR} \; - $ \ model. For ease of viewing, values have been rounded.}
\tablenotetext{f}{VLT combined nights data.}
\label{tab:S0-2_measure}
\end{deluxetable*}

\newpage
\newpage
\appendix

\section{Orbital Fit}	\label{sect:full_orbit_res}
The model for S0-2's long-term RV variation is based on a joint orbital fit of S0-2 and S0-38. We used the same S0-2 and S0-38 astrometry\footnote{We do not report new astrometric measurements, as additional astrometric data is not expected to significantly affect S0-2\vtick s RV curve.} and process as \citet{Boehle:2016ko}, S0-38 RV from \citep{Gillessen:2017fa}, but with the S0-2 RVs from Table \ref{tab:S0-2_measure}. It should also be noted that the impact of S0-38 is negligible for this S0-2 binary study. The resulting orbital parameters are listed in Table \ref{tab:full_orbit_table}, with all results being consistent with \citet{Boehle:2016ko} within 1$\sigma$.

\begin{deluxetable*}{lr}
\tablecolumns{2} 
\tablewidth{0pc} 
\tablecaption{Results from Orbital Fit}
\tablehead{
\colhead{Model Parameter (units)} &
\colhead{Parameter Value\tna}
}
\startdata
\textbf{Black Hole Properties:} 				&						\\
\hspace{1mm}Distance (kpc) 						&	$7.93 \pm 0.13 \pm 0.04$		\\
\hspace{1mm}Mass ($10^{6}$ \ \msun) 			&	$4.03 \pm 0.14 \pm 0.04$		\\
\hspace{1mm}X position of Sgr A* (mas) 			&	$2.17 \pm 0.47 \pm 1.90$		\\
\hspace{1mm}Y position of Sgr A* (mas) 			&	$-4.31 \pm 0.60 \pm 1.23$	\\
\hspace{1mm}X velocity (mas yr$^{-1}$) 			&	$-0.11 \pm 0.03 \pm 0.13$	\\
\hspace{1mm}Y velocity (mas yr$^{-1}$) 			&	$0.67 \pm 0.06 \pm 0.22$		\\
\hspace{1mm}Z velocity (\kms) 					&	$-9.99 \pm 6.25 \pm 4.28$	\\
\textbf{S0-2 Properties:} 					&						\\
\hspace{1mm}Period (yr)							&	$15.92 \pm 0.04$	\\
\hspace{1mm}Time of closest approach (yr)		&	$2018.266 \pm 0.04$\\
\hspace{1mm}Eccentricity						&	$0.892 \pm 0.002$	\\
\hspace{1mm}Inclination (deg)					&	$134.3 \pm 0.3$		\\
\hspace{1mm}Arguement of periapse (deg)			&	$66.7 \pm 0.5$		\\
\hspace{1mm}Angle of the ascending node (deg)	&	$228.0 \pm 0.5$		\\
\textbf{S0-38 Properties:} 					&						\\
\hspace{1mm}Period (yr)							&	$19.20 \pm 0.2$		\\
\hspace{1mm}Time of closest approach (yr)		&	$2003.1 \pm 0.04$	\\
\hspace{1mm}Eccentricity						&	$0.811 \pm 0.004$	\\
\hspace{1mm}Inclination (deg)					&	$170 \pm 2$			\\
\hspace{1mm}Arguement of periapse (deg)			&	$194 \pm 160$		\\
\hspace{1mm}Angle of the ascending node (deg)	&	$79 \pm 24$

\enddata
\tablenotetext{a}{The first error term for each best-fit value corresponds to the statistical error determined by the orbital fit. For the black hole parameters, the second error term corresponds to jackknife uncertainty from the reference frame, which were reported in \citet{Boehle:2016ko}.}
\label{tab:full_orbit_table}
\end{deluxetable*}

\section{Lomb-Scargle Method}	\label{sect:lombscargle}
In this work, we used the Lomb-Scargle method to look for periodic signals in the S0-2 data. The Lomb-Scargle method works best at detecting sinusoidal signals, which corresponds to a circular binary system. However, as binaries become eccentric, their radial velocity curve deviates more from a perfect sine wave. Although their curves are periodic, their non-sinusoidal shapes could lead to reduced sensitivity using the Lomb-Scargle periodogram.

\begin{figure}[htb]
\centering
\includegraphics[width=\linewidth]{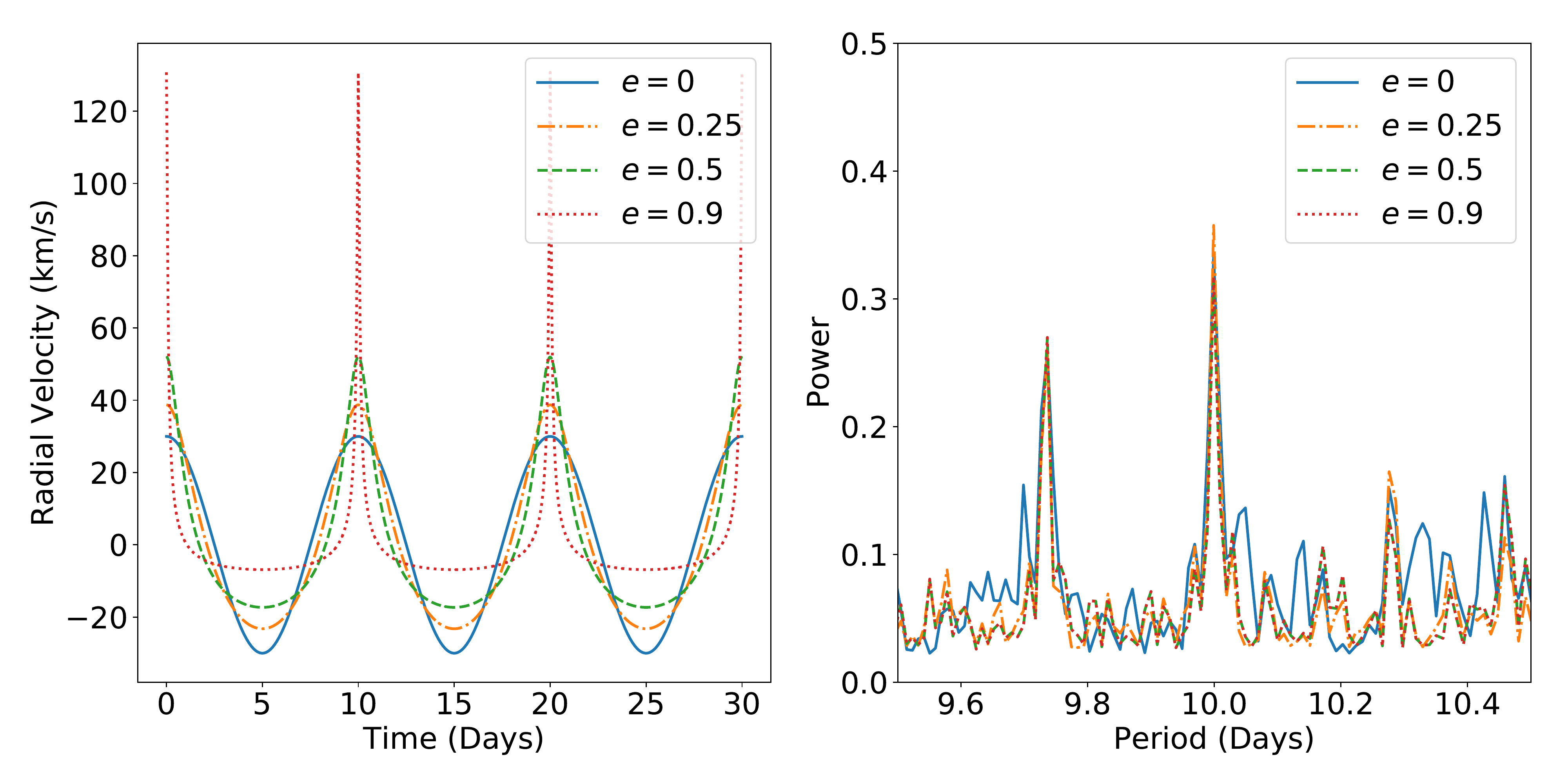}
\caption{
\small
Left: Sample radial velocity curves with different eccentricities. Each curve has a period of 10 days, a $K$ value of 30 \kms and $\omega$ of 0 degrees. Each curve was sampled at the same times as our data. Right: Median Lomb-Scargle power values for each set of 100,000 simulations run for the different eccentric curves.
\label{fig:ls_ecc}
}
\end{figure}

We explored the method\vtick s sensitivity to eccentricity by generating four sets of 100,000 simulated eccentric binary radial velocity curves (see equation \ref{eq:eccentric_rv}). The first set of curves had $e = 0$, the second set had $e = 0.25$, the third set had $e = 0.5$ and the fourth set had $e = 0.9$. All curves had the same period of 10 days, a $K$ value of 30 \kms, and $\omega$ of 0 degrees (Figure \ref{fig:ls_ecc}). These curves were also sampled at the same times as our data. We ran each set of curves through the Lomb-Scargle analysis and took the median power values for each period. The Lomb-Scargle method successfully identified the 10 days period in the different sets of simulated curves. The median power values at the 10 days period varied by less than 0.04 between the sets. We interpret the second peak as a product of the sampling of our observations.

\begin{figure}[htb]
\centering
\includegraphics[width=\linewidth]{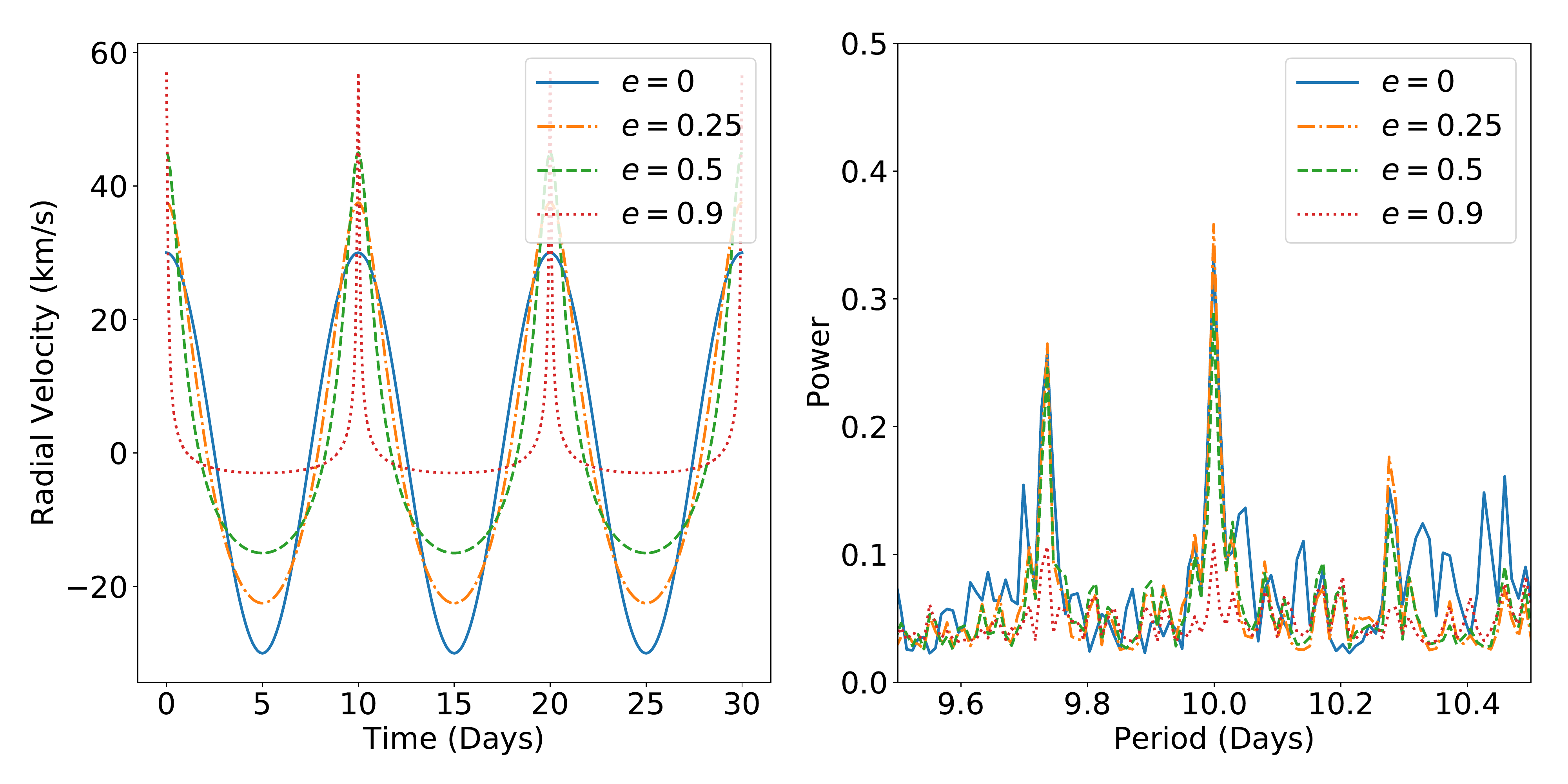}
\caption{
\small
Left: Sample radial velocity curves with different eccentricities. Each curve has a period of 10 days, a RV amplitude of 30 \kms and $\omega$ of 0 degrees. Each curve was sampled at the same times as our data. Right: Median Lomb-Scargle power values for each set of 100,000 simulations run for the different eccentric curves.
\label{fig:ls_ecc_rvamp}
}
\end{figure}

We did this same analysis but instead kept the RV amplitude constant at 30 \kms, where RV amplitude is defined as $(RV_{\text{max}} - RV_{\text{min}})/2$ . To do this, we changed the value of $K$ for each value of eccentricty, which corresponds to changing the semi-major axis since the period remains constant at 10 days (Figure \ref{fig:ls_ecc_rvamp}). We find that the Lomb-Scargle power at 10 days for the different eccentricities differ by less than 0.07 for eccentricities of 0, 0.25, and 0.5. For an eccentricity of 0.9, the power value drops to 0.11 at 10 days. We again interpret the second peak as a product of the sampling of our observations.

\end{document}